**Electron spin transport in a metal-oxide-semiconductor Si two-dimensional inversion channel: Effect of hydrogen annealing on spin scattering mechanism and spin lifetime**

Shoichi Sato[1,2], Masaaki Tanaka[1,2], and Ryosho Nakane[1]

[1]*Department of Electrical Engineering and Information Systems, The University of Tokyo, 7-3-1 Hongo, Bunkyo-ku, Tokyo 113-8656, Japan*
[2]*Center for Spintronics Research Network (CSRN), The University of Tokyo, 7-3-1 Hongo, Bunkyo-ku, Tokyo 113-8656, Japan*

**Abstract**

Electron spin transport in a two-dimensional Si inversion channel was experimentally and theoretically studied in terms of the electron distribution in the subbands, electron momentum scattering processes, electron momentum lifetime, and spin lifetime. The electrical properties, electron charge transport, and spin transport were investigated by measuring a Si-based spin metal-oxide-semiconductor field-effect transistor with a 10 μm channel length under various bias and temperature conditions (4, 77, 150, and 295 K). In particular, in our unique procedure, the same device was measured before and after annealing to quantitatively clarify the change in the spin transport with lower and higher electron mobilities, by excluding device-to-device variability. Even when the distribution of electrons in the subbands, electron mobility, and temperature were significantly changed, the spin flip probability, which is defined as the ratio of electron momentum lifetime to spin lifetime, is nearly constant at ~1/25000 probably due to the Elliot-Yafet mechanism. The spin conservation lengths at various temperatures were increased 2–50 times with the increase in both the electron mobility and spin drift, in which the enhancement factor was found to reasonably agree with a simple theoretical prediction. With a high electron mobility of 3065 $cm^2$/Vs at 4 K, the spin transport with a spin conservation efficiency of 94% was achieved through the 10-$\mu$m-long channel.

**I. Introduction**

Recently, the electron spin transport through a Si channel has generated considerable interest since the spin orbit coupling (SOC) in bulk Si materials can yield electron transport with a long spin conservation length. This character is attractive from the point of view of fundamental material physics and spin-functional device applications [1-9]. Based on the Elliott-Yafet (EY) mechanism that takes unto account the Si band structure and acoustic phonon scatterings [10,11], a spin flip occurs once in some electron momentum scattering events with a constant rate that is proportional to the strength of SOC. Thus, when this mechanism is applicable, the ratio of electron momentum lifetime $\tau$ to the spin lifetime $\tau_S$ is constant, namely, the spin-flip probability (SFP) $\tau/\tau_S$ = constant. An additional spin scattering caused by the imperfection of the lattice, such as impurities and defects, was found to follow the EY mechanism [12]. This is regarded as an extrinsic property of Si. Experimental and theoretical studies on $n$-type bulk Si materials revealed that $\tau_S$ changed with dopant species and their volume concentrations. For instance, arsenic reduces $\tau_S$ by one order of magnitude than phosphorous under a similar concentration [13,14]. On the other hand, a recent theory on optical phonon scatterings predicts that the inter-valley $g$ and $f$ processes are the spin-conserved and spin-flip scatterings, respectively [15]. This is regarded as an intrinsic property of Si and has not yet been experimentally studied. Since it has been unclear whether the SFP differs depending on the electron momentum scattering process or not, it is necessary to clarify the SFP in a Si 2D channel while the dominant electron momentum scattering process is changed.

To clarify the electron spin transport in semiconductors, a very useful platform is a planar-type Si-based spin metal-oxide-semiconductor field-effect transistor (spin MOSFET) [16-18] with ferromagnetic source (S)/drain (D) electrodes and $SiO_2$/Si gate stack. This is because the electron spin transport through the Si two-dimensional (2D) electron channel can be examined while the properties of the electron transport significantly change with the gate electric field $E_G$, lateral electric field $E_L$ parallel to the electron transport, and temperature $T$.

Using $E_G$ and $T$, the electron distribution in the subbands in the 2D electron channel can be changed, which subsequently changes the dominant electron momentum scattering process during the transport. The background of this study and detailed electrical properties of the Si 2D channel are summarized in S.M. [19]. These characteristics offer rich opportunities to systematically study the relation between the electron momentum and spin flip scatterings.

With decreasing $T$ and fixing $E_G$ in the middle range, the electron populations in the subbands with lower energies increases and the contribution of the phonon scattering to the electron momentum scattering is reduced. Using these features, the contribution of the phonon scattering to $\tau/\tau_S$ can be estimated. Moreover, when a high electron mobility $\mu$ value, comparable to that in practical ordinary MOSFETs, is realized, $\tau/\tau_S$ can be estimated when the dominant electron momentum scattering significantly changes in a wide $E_G$ range. Till date, $\tau/\tau_S$ has never been estimated in the lower $E_G$ range where the Coulomb scattering is dominant [8].

On the other hand, the spin conservation length $\lambda_{eff}$ is also important for the establishment of spin-valve device guidelines. When we apply $E_L$, $\lambda_{eff}$ changes while $\tau_S$ remains unchanged under a certain $E_G$ and $T$. This phenomenon varies depending on the dominance of the spin diffusion or spin drift [20]: $\lambda_{eff}$ approaches the spin diffusion length $\lambda_{diff} = \sqrt{D_e \tau_S}$ near the diffusive transport limit with a low $E_L$, whereas $\lambda_{eff}$ approaches the spin drift length $\lambda_{drift} = \mu E_L \tau_S$ near the drift transport limit with a high $E_L$, where $D_e$ and $\mu$ are the electron diffusion coefficient and electron mobility, respectively. The enhancement of $\lambda_{eff}$ by the spin drift in a Si channel has been experimentally verified by other groups [21,22]. Our previous experiments on a Si-based spin MOSFET [8] demonstrated the spin drift effect at room temperature. We observed clear spin signals in a 10-$\mu$m-long 2D inversion channel under a high $E_L$ although $\lambda_{diff}$ at $E = 0$ V/cm was estimated to be ~1 $\mu$m. In consequence, the spin drift was found to be actually effective in the 2D inversion channel in the middle-to-higher $E_G$ range. Since $\lambda_{\text{drift}}$ is proportional to both $\mu$ and $\tau_S$, $\alpha$ times increase in $\tau$ under a constant $\tau/\tau_S$ increases $\lambda_{\text{drift}}$ by $\alpha^2$ times. This theoretical relation expected from the combination of the spin drift and EY mechanism has yet not been experimentally verified.

In this paper, we study the electron spin transport in a Si 2D inversion channel with various $\mu$ values while $E_G$, $T$, and $E_L$ are varied in a wide range to comprehensive understand the spin transport. The main purposes are as follows: (1) to deeply understand the spin transport in a Si channel by estimating $\tau/\tau_S$ when the electronic properties and electron momentum scattering processes are changed, and (2) to estimate the increase in $\lambda_{\text{eff}}$ when $\tau$ is increased $\alpha$ times and examine whether $\lambda_{\text{eff}}$ increases $\alpha^2$ times or not. We measure both as-prepared and annealed spin MOSFET devices with low and high $\mu$ values, respectively, which enables us to quantitatively reveal change in the spin transport with change in the channel properties, by excluding device-to-device variability.

**II. Device structure**

Figure 1(a) shows a schematic of a back-gate-type spin MOSFET on a silicon-on-insulator (SOI) substrate with a 200-nm-thick $SiO_2$ buried oxide (BOX) layer used in this work. In this device, Fe/Mg/MgO/$n^+$-Si ferromagnetic tunnel junctions are used as the S/D, a 8-nm-thick $p$-Si channel with a boron doping concentration $N_A = 1\times10^{15}$ cm$^{-3}$ has a length $L_{\text{ch}}$ = 10 $\mu$m and width $W_{\text{ch}}$ = 180 $\mu$m, the BOX layer is used as the gate dielectric, and reference electrodes L and R are located outside of the S and D, respectively. Note that the channel length $L_{\text{ch}}$ is defined by the distance between the $n^+$-Si regions at the S and D. Figure 1(b) shows the detailed structure of the ferromagnetic tunnel junctions at the S and D, in which the width of the bottom $n^+$-Si layer with a phosphorus doping concentration $N_D = 1.3 \times 10^{20}$ cm$^{-3}$ is the same as the channel width ($W_{ch}$ = 180 $\mu$m) and $L_S$ = 0.7 $\mu$m for the S and $L_D$ = 2.0 $\mu$m for the D. On the same substrate, a bottom-gate-type Hall bar MOSFET with the same channel property and junction structure was also fabricated (see S.M.[19]), where the channel length and width are 460 and 90 $\mu$m, respectively. The detailed fabrication process of the devices is described in Ref. [8]. We examined the electrical properties of the Si 2D inversion channel in two types of spin MOSFET devices (Fig. 1): i) as-prepared device and ii) annealed device; first, the as-prepared device was measured, annealed at 250°C for 30 min in a $N_2$+ 4% $H_2$ ambient, and then the annealed device was measured. Both devices were measured under various bias and temperature conditions. The electron charge and spin transport were measured at $T$ = 4, 77, 150, and 295 K to examine various scattering processes and electron distributions in the subbands. Note that the L and R electrodes are not used in this study.

**III. Electrical properties of the Si 2D inversion channel**

Two methods were used to investigate the electrical properties of the Si 2D inversion channel: a Hall measurement with the Hall bar MOSFET and drain-to-source current $I_{DS}$ versus gate-to-source $V_{GS}$ measurement with the spin MOSFET. Then, the following parameters were estimated: $\mu$, sheet electron density $N_S$, threshold voltage $V_{th}$, and channel sheet resistance $R_S$. Next, the energy level and electron population at each subband were estimated using the experimental results and self-consistent calculation using Schrödinger and Poisson equations [7,8,23]. Here, the effective electron conduction mass $m^*$ was calculated by $m^* = \sum_i \sum_{v=2,4} m^{(v)} N_i^{(v)} / N_S$, where $v$ (= 2 or 4) denotes the $v$-fold subband group, $m^{(v)}$ denotes the effective mass in the $v$-fold subband group ($m^{(2)}$ = 0.19 and $m^{(4)}$ = 0.315), $i$ (= 0, 1, 2,

⋯) denotes the subband number counted from the energy minimum of each subband group, and $N_i^{(v)}$ denotes the sheet electron density at the $i$th energy in the $v$-fold subband group. Finally, $\tau$ values at various $V_{GS}$ and $T$ were estimated using the relation $\tau = m^*\mu/q$, where $q$ is the elemental charge. To characterize the scattering process, $\mu$ and $\tau$ will be plotted as a function of $N_S$.

Figures 2(a) and (b) show $\mu$ versus $N_S$, where open and closed circles are the estimated values of the as-prepared and annealed devices, respectively, each color denotes $T$, and dashed and solid lines connecting nearest-neighbor circles are guides. The $\mu$-$N_S$ curve for the as-prepared device (Fig. 2(a)) changes little with a decrease in $T$, which indicates that the temperature-insensitive Coulomb and surface roughness scattering processes govern the electron charge transport. In contrast, the $\mu$-$N_S$ curve of the annealed device (Fig. 2(b)) increases significantly with decreasing $T$, which indicates the decrease in the temperature-sensitive phonon scattering processes. In these figures, the scattering processes that determine the electron mobility are plotted using the well-accepted formulae, where black solid, dotted, and bold lines denote the Coulomb scattering ($\propto N_S^{+1}$), phonon scattering at 295 K ($\propto N_S^{-0.3}$), and surface roughness scattering ($\propto N_S^{-2}$), respectively [24]. The Coulomb, phonon, and roughness scatterings were confirmed to be dominant in the lower, middle, and higher $N_S$ ranges at 295 K, respectively, and the Coulomb and roughness scatterings are dominant in the lower and higher $N_S$ ranges at 4 K, respectively. The effect of the annealing on the $\mu$-$N_S$ curve shape can be clearly observed in the lower $N_S$ range. With decreasing $N_S$, the $\mu$ curve for the as-prepared device bends towards the bottom at $\sim 5 \times 10^{12}$ cm$^{-2}$, whereas that for the annealed device bends at $\sim 1.5 \times 10^{12}$ cm$^{-2}$. Since the Coulomb scattering is the dominant scattering process in the lower $N_S$ range and it causes the $\mu$-$N_S$ curve to bend, the significant increase in $\mu$ by the annealing is attributable to the reduction in the number of the scattering events by the Coulomb scattering, which is shown by the black solid lines in Figs. 2(a) and (b). It is most probable that hydrogen termination by annealing deactivates fixed charges either at the Si/SiO$_2$ interface or in the SiO$_2$ layer, leading to the reduction in the fixed defect state density [25]. This conclusion is supported by the experimental result where $V_{th}$ negatively shifted from 13 to 2 V due to the annealing (see, S.M.[19]). It is noteworthy that the maximum $\mu$ = 3065 cm$^2$/Vs at 4 K is comparable to that (~ 4000 cm$^2$/Vs) at 25 K of a practical SOI MOSFET [26]. This clearly proves that our annealed device has a high-quality MOS gate stack with a low fixed defect state density.

Figures 2(c) and (d) show $N_i^{(v)}$ versus $N_S$ plots calculated at 4 and 295 K, respectively, where red and pale red regions denote the electron populations at the 2-fold subband with the lowest ($i$ = 0) and higher ($i$ = 1–9) energy levels, respectively, and blue and pale blue regions denote the electron populations at the 4-fold subband with the lowest ($i$ = 0) and higher ($i$ = 1–9) energy levels, respectively. The detailed calculation method is described in S.M. of refs [7, 8]. Here, we used $N_A$ = 10$^{15}$ cm$^{-3}$, $t_{Si}$ = 8 nm, and other parameters taken from ref. [23]. Below $N_S$ = 2×10$^{12}$ cm$^{-2}$, all the electrons populated at the bottom energy level of the 2-fold subband at 4 K ($N_0^{(2)}$ = 100%), whereas electrons distributed in various energy levels in the 2-fold and 4-fold subband groups at 295 K. Figure 2(e) shows $N_i^{(v)}$ versus $T$ plot calculated for $N_S$ = 2.3 × 10$^{12}$ cm$^{-2}$, where $\mu$ at 4 K of the annealed device shows the maximum value. Considering Figs. 2(a) and (b), the electron momentum scattering for the $N_S$ value in Fig. 2(e) is as follows. In the as-prepared device, the Coulomb scattering is the dominant at all the values of $T$. In the annealed device, the Coulomb scattering is dominant at 4 K, the intra- and inter-valley phonon scatterings contribute more with increase in $T$, and finally the intra- and inter-valley phonon scatterings are dominant at 295 K. For detailed analysis of the temperature dependence, we calculated the phonon scattering rates in various temperature using our self-consistent calculation (see S.M. for the details [19]). We confirmed that the contribution of the intervalley *f*-process shows largest increase among the phonon scatterings with an increase in the temperature; the contribution of the inter-valley *f*- and *g*-processes steeply increases with an increase in the temperature, and the former is more than four times greater than the latter above 200 K. Thus, we can mainly evaluate the contribution of the following scattering processes to the spin flip:

A) The Coulomb scattering

Since the Coulomb scattering dominates at 4 K in both devices and annealing reduces the scattering rate, the contribution of the Coulomb scattering can be evaluated by comparing the increase of $\tau$ and $\tau_S$ at 4 K.

B) The *f*-process inter-valley phonon scattering

In the annealed device, the contribution of the intervalley *f*-process increased the most with an increase in the temperature. Therefore, SFP due to the intervalley *f*-process can be mainly evaluated when the temperature dependences of $\tau$ and $\tau_S$ are analyzed in the annealed device.

**IV. Electron spin transport**

To estimate $\tau_S$ in the Si 2D inversion channel, two-terminal Hanle spin precession signals (2TH signals) [27,28] were measured with the setup shown in Fig. 1(a). The voltage between the S and D was measured with a constant $V_{GS}$ and D-to-S current $I_{DS}$ (= 5, 8, and 10 mA) while a perpendicular external magnetic field $H_\perp$ was applied and swept between ±3000 Oe. For measurements at various *T*, the $V_{GS}$ value corresponding to the maximum $\mu$ was used; at such $V_G$, $N_S$ = 3.1×10$^{12}$ cm$^{-2}$ for the as-prepared device and $N_S$ = 2.0×10$^{12}$ cm$^{-2}$ for the annealed device. Under these $N_S$ conditions, the electron distribution was calculated as shown in Fig. 2(e). In addition, 2TH signals were measured with various $N_S$, as shown in Table 1.

Prior to the measurement of 2TH signals, the relative magnetization configuration of the S and D electrodes was set at the parallel (P) or antiparallel (AP) magnetization state using the major or minor loop of spin-valve signals with an in-plane external magnetic field. It is well known that voltage signals experimentally measured using the setup in Fig. 1(a) include both the spin transport and local spin accumulation signals [8,29,30,31]. In this study, we used the following formula to extract the spin transport signal, $\Delta V^{2TH}(H_\perp) = [V^{AP}(H_\perp) - V^{P}(H_\perp)]/2$, where $V^{AP}$ and $V^{P}$ denote the voltage signals obtained in the AP and P magnetization states, respectively.

Figures 3(a) and (b) show 2TH signals at 4 K measured with $I_{DS}$ = 5, 8, and 10 mA for the as-prepared and annealed devices, respectively, where red curves are experimental signals. In the 2TH signals, the oscillation and envelope originate from the in-phase and dephasing of the spin precession, respectively [1]. Hence, the clear Hanle precession curves in Figs. 3(a) and (b) represent the electron spin transport in both the as-prepared and annealed devices under all the bias conditions. In the same manner, 2TH signals were measured at various $T$. Figures 3(c) and (d) show Hanle precession curves measured at $T$ (= 4, 77, 150, and 295 K) with $I_{DS}$ = 10 mA for the as-prepared and annealed devices, respectively, where red curves are experimental signals. In both cases, the amplitude of the signal increases with decrease in $T$, indicating that the spin polarization of electrons increases with decreasing $T$. When the signals at the same $T$ of the both devices are compared with each other, the number of the oscillation for the annealed device is larger than that for the as-prepared device. The increase in this number by the annealing indicates the suppression of spin dephasing, namely, the spin drift effect largely contributes to the spin transport in the annealed device than in the as-prepared device. On the other hand, in each device, the magnetic field positions for the peak/valley in the oscillation remain unchanged when $T$ is changed. This is clear that the drift velocity $v_d = \mu E_L$ remain constant with constant $N_S$ and $I_{DS}$ (= $qN_S v_d W_{ch}$), although $\mu$ drastically changes ($\mu$ = 265–441 cm$^2$/Vs in the as-prepared device and $\mu$ = 451–3065 cm$^2$/Vs in the annealed device) when $T$ was changed from 295 to 4 K.

Next, all the 2TH signals were fitted using the following equation that takes into account the spin drift effect [8,20]:

$$\Delta V^{2TH(P/AP)}(H_\perp) = -\sigma^{P/AP} \text{Re}\left[ \frac{P_S^2 I_{DS}}{X} \left( \frac{1}{r_{ch}^u} + \frac{1}{r_{ch}^d} \right) \exp\left( -\frac{L_{ch}}{\lambda_{ch}^d} \right) \right], \quad (1)$$

$$X = \left( \frac{1}{r_{NL}^{(S)}} + \frac{1}{r_{ch}^d} \right)\left( \frac{1}{r_{NL}^{(D)}} + \frac{1}{r_{ch}^u} \right) - \left( \frac{1}{r_{NL}^{(S)}} - \frac{1}{r_{ch}^u} \right)\left( \frac{1}{r_{NL}^{(D)}} - \frac{1}{r_{ch}^d} \right)\exp\left( -\frac{L_{ch}}{\lambda_{ch}^u} - \frac{L_{ch}}{\lambda_{ch}^d} \right), (2)$$

$$\frac{1}{\hat{\lambda}_{ch}^{d(u)}} = -\frac{\mu E_L}{2D_e} + \sqrt{\left(\frac{\mu E_L}{2D_e}\right)^2 + \left(\frac{1}{\hat{\lambda}_{ch}}\right)^2}, \quad (3)$$

$$r_{NL}^{S(D)} = \frac{r_{ch} + r_n \tanh(L_{S(D)}/\lambda_n)}{r_n + r_{ch} \tanh(L_{S(D)}/\lambda_n)} r_n, \quad (4)$$

where $\sigma^{P/AP} = +1/-1$ is the sign factor for the P/AP magnetization configuration, $P_S$ is the tunneling spin polarization of the Fe/Mg/MgO/$n^+$-Si junction at the S and D electrodes, $r_{ch} = R_S \hat{\lambda}_{ch}/W_{ch}$ is the intrinsic spin resistance in the inversion channel, $r_{ch}^{u(d)} = R_S \hat{\lambda}_{ch}^{d(u)}/W_{ch}$ is the effective spin resistance with the spin drift in the inversion channel, $r_n = \rho_n \hat{\lambda}_n / t_n W_{ch}$ is the spin resistance in the $n^+$-Si regions below the S and D electrodes, $L_S$ = 0.7 $\mu$m and $L_D$ = 2.0 $\mu$m are the $n^+$-Si region lengths parallel to the electron transport direction, $\hat{\lambda}_n = \sqrt{D_n \tau_n/(1+i\gamma H_\perp \tau_n)}$ and $\hat{\lambda}_{ch} = \sqrt{D_e \tau_S/(1+i\gamma H_\perp \tau_S)}$ are the complex spin diffusion lengths [32] in the $n^+$-Si regions and inversion channel, respectively, and $\tau_n$ and $D_n$ are the spin lifetime and electron diffusion coefficient in the $n^+$-Si regions, respectively. The effective spin resistance of the non-local regions $r_{NL}^{S(D)}$ is the series spin resistance composed of $r_n$ and $r_{ch}$. In the fitting, the experimentally estimated $\tau_n$ = 0.6–0.8 ns and $D_n$ = 3.5–4.4 cm$^2$/s were used, as in our previous paper [8]. From Eq. (1), it is clear that the equation of $\Delta V^{2TH(P/AP)}$ is a nonlinear function of the various fitting parameters ($P_S$, $\tau_S$, $\mu$, and $D_e$). For highly accurate estimation, the signals with $I_{DS}$ = 5, 8, and 10 mA are fitted so that all the fittings converged with the same $D_e$, $\mu$, and $\tau_S$ that are independent of $I_{DS}$. As shown in Figs. 3(a)–(d), the fitting curves (black dashed curves) almost perfectly reproduce all the experimental signals, from which we obtained $\tau_S$ = 1.4 ns and $\mu$ = 410 cm$^2$/Vs for the as-prepared device and $\tau_S$ = 9.3 ns and $\mu$ = 3074 cm$^2$/Vs for the annealed device. The validity of the fittings is supported by the fact that the $\mu$ values estimated from the 2TH signals are almost identical with those estimated from the electrical measurements, as shown in Figs. 4(a) and (b). In the same manner, the 2TH signals measured at various temperatures were fitted by Eq. (1), as shown by black dashed curves in Figs. 3(c) and (d). As in the case of 4 K, the experimental and fitting curves agree very well with each other.

## V. Spin flip probability and the spin conservation length

Figures 5(a) and (b) show $\tau_S$ and $\tau$ plotted as a function of $N_S$ for the as-prepared and annealed devices, respectively, and Fig. 5(c) shows $\tau_S$ and $\tau$ plotted as a function of $T$, where squares are $\tau_S$ estimated from the 2TH signals, circles are $\tau$ estimated from the electrical measurements (Figs. 2(a) and (b)), and left and right vertical axes represent the scales for $\tau_S$ and $\tau$, respectively. Note that the scale of the left axis is 25000 times larger than that of the right axis in Fig. 5 (a)–(c). As described in Section III, we discuss the contributions of the A) Coulomb scattering and B) inter-valley *f*-process phonon scattering to the spin flip based on our experimental results and theoretical fittings.

A) The Coulomb scattering

As shown by the blue symbols in Figs. 5(a) and (b), both $\tau_S$ and $\tau$ at 4 K increased ~6 times by the $H_2$ annealing while the ratio $\tau/\tau_S$ (~1/25000) remains unchanged, indicating that SFP of the Coulomb scattering is ~1/25000.

B) The *f*-process inter-valley phonon scattering

As shown by the filled symbols in Fig. 5(c), $\tau_S$ and $\tau$ in the annealed device have almost the same temperature dependences, indicating that the SFP of the *f*-process inter-valley scattering is ~1/25000.

From the abovementioned results, we found $\tau/\tau_S$ is almost constant at ~1/25000 under any conditions, i.e., the EY mechanism dominates SFP ($=\tau/\tau_S$) in the Si 2D inversion channel even when the electrical properties and dominant scattering process changed significantly. The EY mechanism is supported by the increased $\tau_S$ with increase in $\mu$ by the reduction in the fixed defect state density due to annealing. This is because the anti-correlation between $\tau_S$ and $\mu$ should be obtained when the D'yakonov-Perel' (DP) process is dominant [33,34]. It should be noted that the $\tau/\tau_S$ value (~1/25000) in the *p*-Si channel with lightly boron doping ($N_A = 1 \times 10^{15}$ cm$^{-3}$) is smaller than that (~1/14000) in the phosphorous-doped 2D accumulation channel ($N_D = 1 \times 10^{17}$ cm$^{-3}$) when $T = 295$ K [8]. Considering that ionized dopant atoms can also be spin-flip scattering centers at 295 K, they likely originate from the differences in the channel properties, such as the doping concentration and dopant atoms.

From Refs [20], $\lambda_{\text{eff}}$ is given by

$$\lambda_{\text{eff}} = \left[-\frac{\mu E_L}{2D_e} + \sqrt{\left(\frac{\mu E_L}{2D_e}\right)^2 + \left(\frac{1}{\lambda_{\text{diff}}}\right)^2}\right]^{-1} = \frac{\lambda_{\text{drift}} + \sqrt{\lambda_{\text{drift}}^2 + 4\lambda_{\text{diff}}^2}}{2}, \quad (5)$$

where $\lambda_{\text{diff}} = \sqrt{D_e \tau_S}$ and $\lambda_{\text{drift}} = \mu E_L \tau_S$. To further understand the spin transport physics, it is worth plotting the experimental $\lambda_{\text{eff}}$ and its extended curves calculated by Eq. (5) for a wide range of $E_L$. Figure 5(d) shows $\lambda_{\text{eff}}$ plotted as a function of $E_L$, where open and closed circles are the experimental values estimated for the as-prepared and annealed devices, respectively, and dashed and solid lines are the extended curves for the as-prepared and annealed devices, respectively. When $E_L$ increased, the dashed curves for the as-prepared device show a gradual increase at first, followed by a drastic increase. The former and latter increases correspond to the diffusion-dominated $\lambda_{\text{eff}}$ ($\sim \lambda_{\text{diff}}$) and drift-dominated $\lambda_{\text{eff}}$ ($\sim \lambda_{\text{drift}}$), respectively. Regarding the solid curves for the annealed device, the $\lambda_{\text{eff}}$-$E_L$ curves at 4 and 77 K shows a monotonic increase due to the drift-dominated $\lambda_{\text{eff}}$ in the entire $E_L$ range, which is confirmed from the fact that these lines are parallel to the drift-dominated $\lambda_{\text{eff}}$ for the as-prepared device. From these features, it can be concluded that all the experimental results denoted by the closed and open circles are $\lambda_{\text{eff}}$ enhanced by the spin drift. From Figs. 5(a)-(d), we confirmed the following theoretical relation: the drift-dominated $\lambda_{\text{eff}}$ ($\sim\lambda_{\text{drift}}$) increases $\alpha^2$ times when $\tau$ is increased $\alpha$ times [See section S3 in S.M. [19] for the details. As shown in Fig. S4, in the entire $T$ range (4–295 K), $\beta$ and $\zeta$ agree with $\alpha^2$ and $\alpha$, respectively, where $\beta = \lambda_{\text{eff}}^{\text{ANL}}/\lambda_{\text{eff}}^{\text{AS}}$ and $\zeta = \tau_S^{\text{ANL}}/\tau_S^{\text{AS}}$, and the superscripts AS and ANL denote the values for the as-prepared and annealed devices, respectively.]

For deeper understanding of the electron spin transport in the Si 2D inversion channel, it is also important to analyze $\lambda_{\text{eff}}$ and the spin conservation efficiency $\eta_S$ during the transport through the 10-$\mu$m-long inversion channel. From the experimental values at 4 K in Fig. 5 (d), the as-prepared and annealed devices have $\lambda_{\text{eff}}$ = 3.9 and 159 $\mu$m at $E_L$ = 556 V/cm, respectively, where the $E_L$ value corresponds to $I_{\text{DS}}$ = 10 mA in the annealed device. Using these values and $\eta_S = \exp(-L_{ch}/\lambda_{eff})$ with $L_{ch}$ = 10 μm, $\eta_S$ = 8% in the as-prepared device and $\eta_S$ = 94% in the annealed device, indicates that $\eta_S$ was dramatically enhanced by the annealing.

## VI. Discussion

Our experiments and analyses revealed that the spin MOSFETs with the high/low $\mu$ values have the constant SFP (= $\tau/\tau_S$) ~1/25000 under any bias current and $T$ conditions, namely, SFP is determined by the EY mechanism even when the channel properties were largely changed.

So far, the changes in $\tau_S$ in a Si 2D channel with temperature have not been clarified, to the best of our knowledge. Thus, it is worth noting that the change in $\tau_S$ in Fig. 5(c) is less than 10 times in the $T$ range of 4–295 K, which is more gradual than that of a theoretical model and experimental results in bulk Si. When only the phonon scattering in Si bulk materials is considered, a theory predicts that the $\tau_S$ - $T$ relation follows a $T^{-5/2}$ law [11], whereas electron spin resonance and spin transport experiments revealed that the $\tau_S$ - $T$ relation follows a $T^{-3}$ law [31]. This is likely because, in Si 2D channels, the Coulomb scattering is dominant at low temperatures, that is, it limits $\tau$ where the strength of phonon scattering is weak (see section S1 in S.M.[19]). On the other hand, at 295 K, $\tau_S$ = 1.9 ns for the annealed device, as shown in Fig. 5(c), which is significantly smaller than $\tau_S$ ~8 ns in the experimental results of Si bulk materials [35]. This is probably due to the quantum confinement of the channel. In Si 2D channels, the strength of phonon scattering is enhanced by a form factor, determined by both envelope wavefunctions of the initial and final states in the scattering process [26,36]. Our self-consistent calculation revealed that the phonon scattering strength is enhanced approximately thrice in our device with an 8-nm-thick channel [19], which reasonably explains $\tau_S$ = 1.9 ns for the annealed device. The above consideration is not applied to the as-prepared device because the Coulomb scattering is dominant at any temperatures, as shown in Fig. 5(c).

The detailed $\tau/\tau_S$ values at different temperatures may explain the physics of the electron spin transport. In the plot for the annealed device in Fig. 5(c), there are differences between the open and closed circles, particularly at 4 and 77 K. The $\tau/\tau_S$ value at 4 K is larger by 16% than 1/25000, whereas that at 77 K is smaller by 11% than 1/25000. We consider that the deviation indicates the difference in SFP between the dominant scattering processes. Nevertheless, SFP remains nearly unchanged depending on the momentum scattering process.

Here, we would like to briefly overview SFP values reported in other materials where the EY mechanism through phonon scattering processes is dominant. As summarized in refs. [37], SFP in Al and Cu are estimated to be ~1/10000 and ~1/1000, respectively, by several experiments on bulk materials, films, and nanowires as well as theoretical studies. On the other hand, SFP in a 85-nm-thick $n^+$-Si with a phosphorus doping concentration of ~$5\times10^{19}$ cm$^{-3}$ was estimated to be ~1/100000 by a spin transport experiment [38]. From the general knowledge that larger atomic number $Z$ leads to larger SOC [39], the order of the SFP values in Al, Cu, and the Si 2D channel can be simply understood: Al ($Z = 13$), Cu ($Z = 29$), and Si ($Z = 14$). Despite the single-crystalline Si with the same dopant atoms in both cases, the SFP value in the Si 2D channel is four times of that in the $n^+$-Si channel. To the best of our knowledge, although the reason for the difference is yet unclear, SOC in the 2D channel can become larger than that in Si bulk materials. A possible origin for the increase of SOC is the spatial symmetry breaking along the vertical direction in the 2D channel, as suggested in our previous paper [8].

Next, we focus on the electron spin scattering in the annealed device at 4 K, where the Coulomb scattering is dominant, all the electrons are populated at the bottom energy level of the 2-fold subband, the dopant boron accepter atoms are freeze-out, and the fixed defect states are localized at the Si/SiO$_2$ interface and in the SiO$_2$ layer. Under such circumstances, the electron momentum scattering occurs through the intra-valley scatterings owing to the fixed defect states in the SiO$_2$ layer [40, 41]. On the other hand, although a theory on the impurity-dominated electron spin-flip scattering in Si 2D inversion channels was reported [33], it cannot be applied to our analysis because it assumes that dopant atoms are ionized and the electron spin flip occurs through the inter-valley $f$ process between the 2-fold and 4-fold subbands even at low temperatures. Besides, as pointed out in ref.[33], the spin flip due to the Coulomb scattering through fixed defect states near a SiO$_2$/Si interface has not been clarified. To qualitatively analyze the Coulomb scattering, the strength of SOC was estimated using the theory in ref. [12]. From the difference in the threshold gate voltage $V_{th}$ between the theoretical and experimental values in the annealed device, the fixed defect states density $N_{DEF}$ was estimated to be $2.3 \times 10^{11}$ cm$^{-2}$ [19] that corresponds to the volume density $n_{DEF} = 1 \times 10^{17}$ cm$^{-3}$, according to the same conversion procedure from the sheet to volume densities in ref. [33]. Using this $n_{DEF}$ value and $\tau_S$ = 9.3 ns, the theory in ref. [12] predicts that the fixed defect states in SiO$_2$ have an SOC effectively 2.3 times larger than that for Sb dopant atoms. Thus, another theory is needed to completely clarify the spin flip scattering due to the Coulomb scattering caused by defects. We speculate that spin at Si$^{3+}$ in our SiO$_2$ layer and screening potentials for ionized defects may have an influence on the electron spin scattering [12,42].

Our finding leads to a simple guideline that $\tau_S$ at room temperature is enhanced by the enhancement of $\mu$. This conclusion is also supported by our result that even the highest $\mu$ value of 3065 cm$^2$/Vs at 4 K is not high enough for the appearance of the DP mechanism [33]. Hence, it is important to note that the established Si technology for the enhancement of $\mu$ in ordinary MOSFETs is also useful for spin MOSFETs, as we demonstrated that the SiO$_2$/Si interface with the lower $N_{DEF}$ achieved by annealing leads to the higher $\tau_S$ value at a certain $T$ than that with the higher $N_{\text{DEF}}$. In Fig. 2(b), although the higher $\mu$ values were realized by the annealing, they are slightly lower than the universal mobility of ordinary MOSFETs with a bulk Si substrate. Since this arises from the enhancement of phonon scattering due to the channel confinement [26], higher $\tau_S$ can be obtained using a thicker SOI channel layer or a top-gated spin MOSFET with a bulk Si substrate. On the other hand, since $\tau_S$ increases with the decrease in the temperature in the annealed device, as shown in Fig. 5(c), the suppression of the inter-valley scatterings is effectively increases $\tau_S$ through the increase in the population at the lowest-energy 2-fold subband, as theoretically predicted in ref. [33].

At present, the factor governing the $\tau/\tau_S$ value is still an open question, particularly there is still lack of the theory on the spin flip scattering due to the Coulomb scattering. Therefore, further experimental and theoretical studies are needed to accumulate the knowledge on the physics of electron spin transport in Si 2D inversion channels.

## IV. Summary

We studied the electron charge and spin transport properties in the 10-$\mu$m-long Si 2D inversion channels using the spin MOSFET with the low/high electron mobility $\mu$ values. The electron momentum lifetime $\tau$ and spin lifetime $\tau_S$ were estimated from the electrical characteristics and 2T Hanle precession signals, respectively, under various bias and temperature ($T$ = 4, 77, 150, and 295 K) conditions. The spin flip probability $\tau/\tau_S$ for both different $\mu$ values was found to be ~1/25000 under any bias and temperature conditions, indicating that it is governed by the EY mechanism even when the dominant electron momentum scattering process was largely changed. Owing to the almost constant $\tau/\tau_S$, the spin conservation length $\lambda_{\text{eff}}$ increases with increasing $\mu$ following our predicted relationship. In the spin MOSFET with the high $\mu$ value, the nearly-ideal spin conservation efficiency 94% was achieved at 4 K where the intra-valley scattering due to the Coulomb scattering is dominant. This result indicates that the suppression of the inter-valley scatterings is the key for higher $\tau_S$ and $\lambda_{\text{eff}}$.

**Acknowledgments**

This work was supported in part by Grants-in-Aid for Scientific Research (Nos. 18H05345, 20H05650), CREST of JST (No. JPMJCR1777), and Spintronics Research Network of Japan.

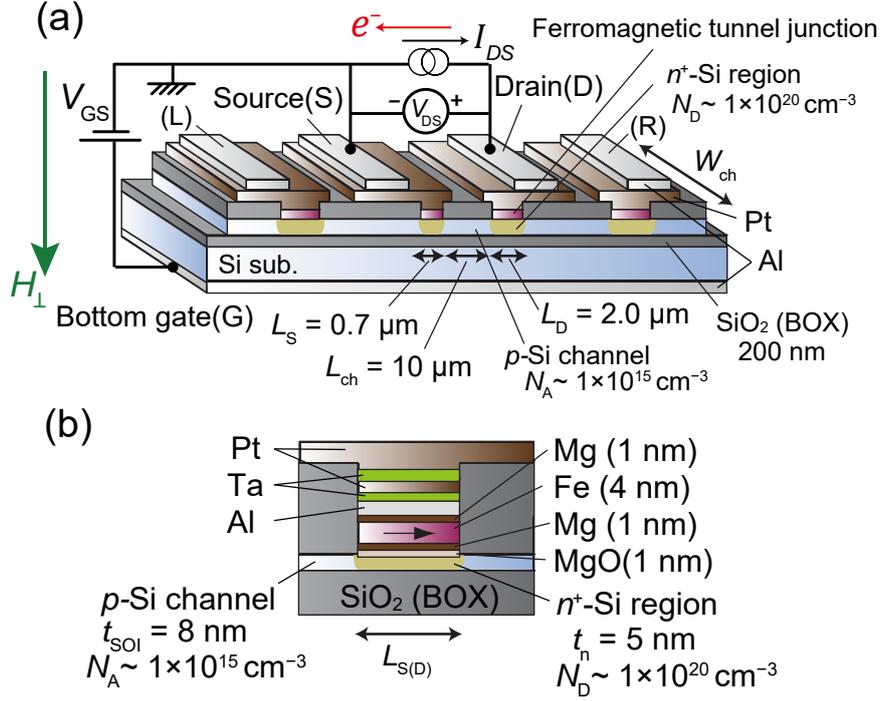

Figure 1 (a) Schematic illustration of a spin metal-oxide-semiconductor field-effect transistor (spin MOSFET) examined in this work. This device was fabricated on a silicon-on-insulator (SOI) substrate with a 200-nm-thick $SiO_2$ buried oxide (BOX) layer. All the top electrodes (S, D, L, and R) were Fe/Mg/MgO/$n^+$-Si ferromagnetic tunnel junctions with a lateral length $L_S = 0.7$ μm for the S and $L_D = 2.0$ μm for the D. A 8-nm-thick $p$-Si channel with a boron doping concentration $N_A = 1\times10^{15}$ cm$^{-3}$ has a length $L_{ch} = 10$ μm and a width $W_{ch} = 180$ μm. The BOX layer is used as the gate dielectric, and reference electrodes L and R are located outside of the S and D, respectively. The measurement setup for two-terminal Hanle (2TH) signals is also illustrated: the voltage between the S and D is measured with a constant drain-to-source current $I_{DS}$ and a constant gate-to-source voltage $V_{GS}$ while an external perpendicular magnetic field $H_\perp$ is swept along the direction parallel to the green allow. (b) Close up view of the Fe/Mg/MgO/$n^+$-Si ferromagnetic tunnel junctions at the top electrodes (S, D, L, and R), where the $n^+$-Si with a phosphorus doping concentration $N_D = 1\times10^{20}$ cm$^{-3}$ has a thickness $t_n = 5$ nm.

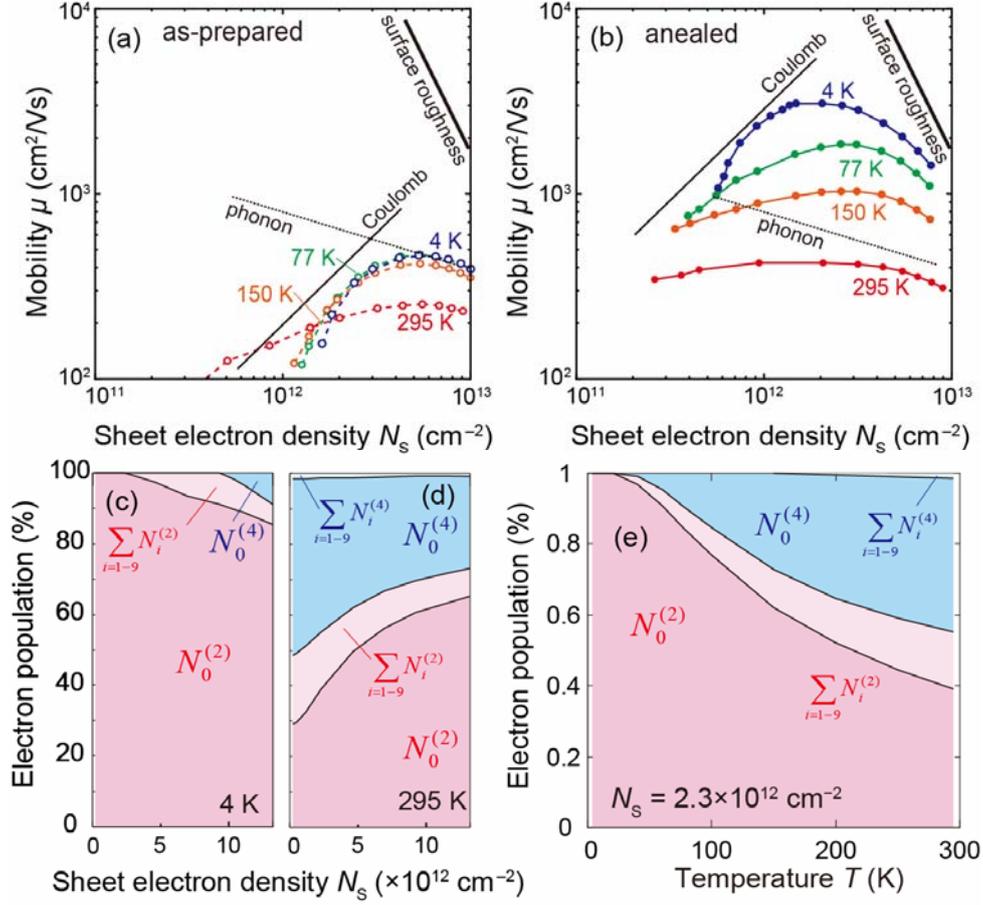

Figure 2 (a)(b) Electron mobility $\mu$ versus the sheet electron density $N_S$ for the (a) as-prepared (open circles) and (b) annealed (closed circles) devices estimated from the electrical measurements at temperature $T$, where blue, green, orange, and red colors denote $T$ = 4, 77, 150, and 295 K, respectively, and lines connecting nearest-neighbor circles are guides. The dominant scattering processes in $\mu$ are also estimated using the simple power law of $N_S$ [19], where black solid, dotted, and bold lines denote the Coulomb ($\propto N_S^{+1}$), phonon ($\propto N_S^{-0.3}$) and surface roughness scatterings ($\propto N_S^{-2}$), respectively. (c)(d)(e) Electron population at each subband in the two-dimensional Si channel at (c) 4 K and (d) 295 K versus $N_S$, and (e) at $N_S$ = 2.3×10$^{12}$ cm$^{-2}$ versus $T$, which were obtained from the electrical properties estimated from the experiments and a self-consistent calculation with Schrödinger and Poisson equations (see SM of ref. [7]). We used $N_A$ = 10$^{15}$ cm$^{-3}$ and $t_{Si}$ = 8 nm, and other parameters are taken from ref. [22]).

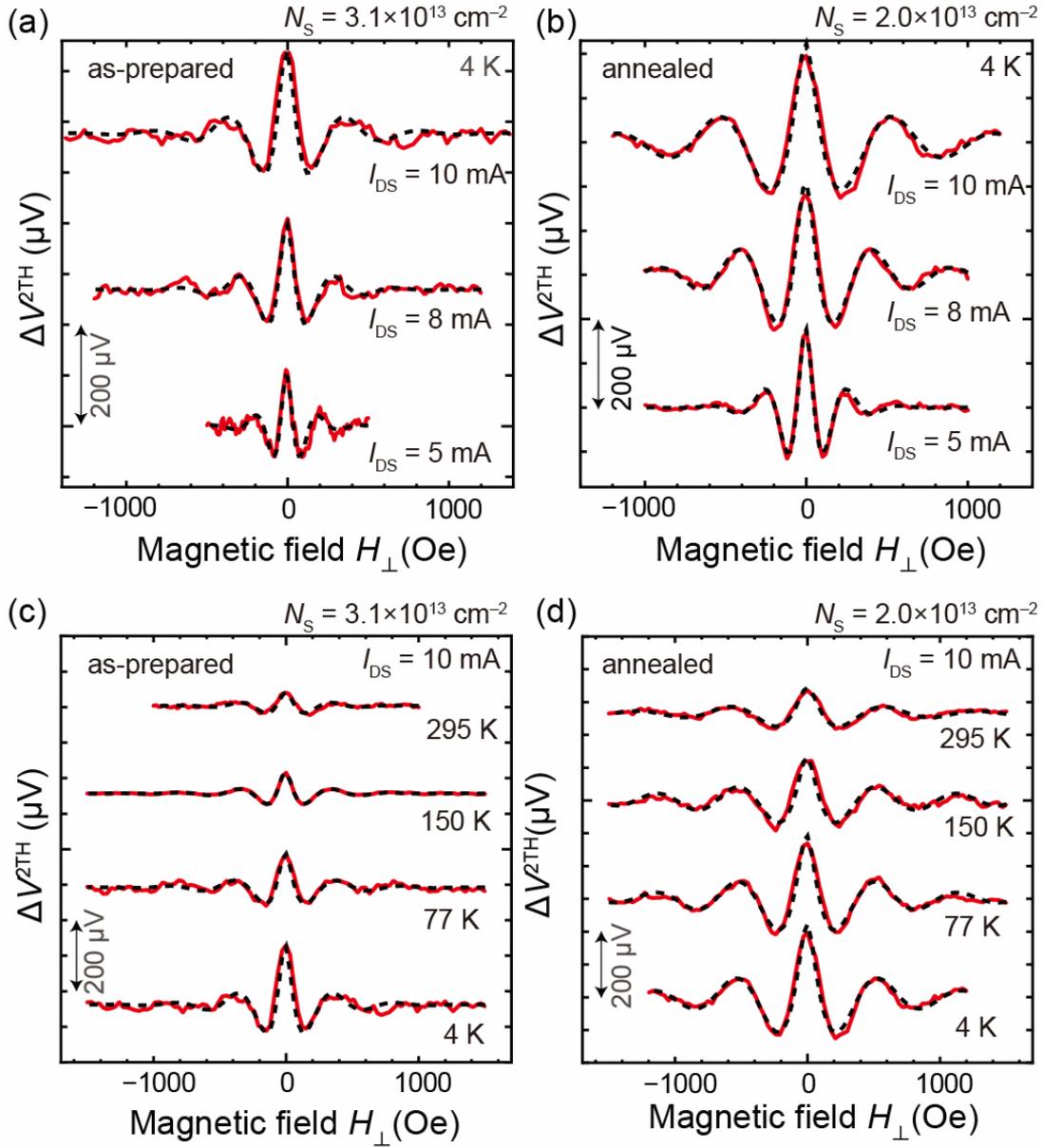

Figure 3 (a)(b) Two-terminal Hanle signals $\Delta V^{2TH}$ measured at 4 K with $I_{DS}$ = 5, 8, and 10 mA for the (a) as-prepared device and (b) annealed device, and (c)(d) $\Delta V^{2TH}$ measured at various temperatures ($T$ = 4, 77, 150, and 295 K) with $I_{DS}$ = 10 mA for the (c) as-prepared device and (d) annealed device, where $N_S$ in (a)(c) and (b)(d) are $3.1 \times 10^{12}$ and $2.0 \times 10^{12}$ cm$^{-2}$, respectively. Red solid and black dashed curves represent the measured signals and fitting results, respectively.  Each curve is vertically shifted for clear vision.

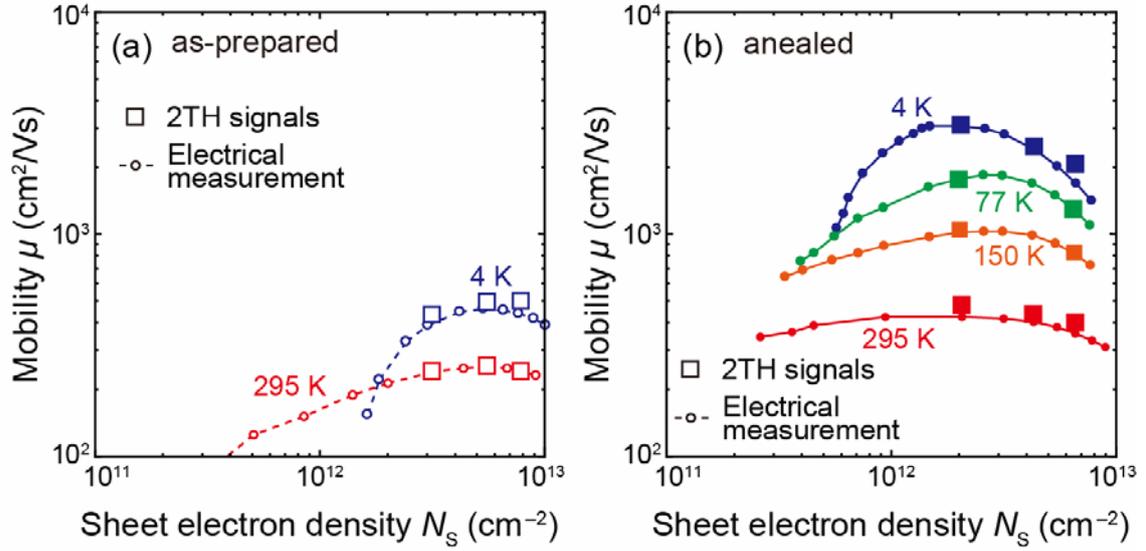

Figure 4 (a)(b) Electron mobility $\mu$ versus the sheet electron density $N_S$, which were estimated for the (a) as-prepared device and (b) annealed device. (a) Open squares are the $\mu$ values estimated from the 2TH signals and open circles with a dashed line are the $\mu$ values estimated from the electrical measurements (Fig. 2(a)). Blue and red colors denote $T = 4$ and 295 K, respectively. (b) Closed squares are the $\mu$ values estimated from the 2TH signals, and closed circles with a solid line are the $\mu$ values estimated from the electrical measurements (Fig. 2(b)). Blue, green, orange, and red colors denote $T = 4$, 77, 150, and 295 K, respectively.

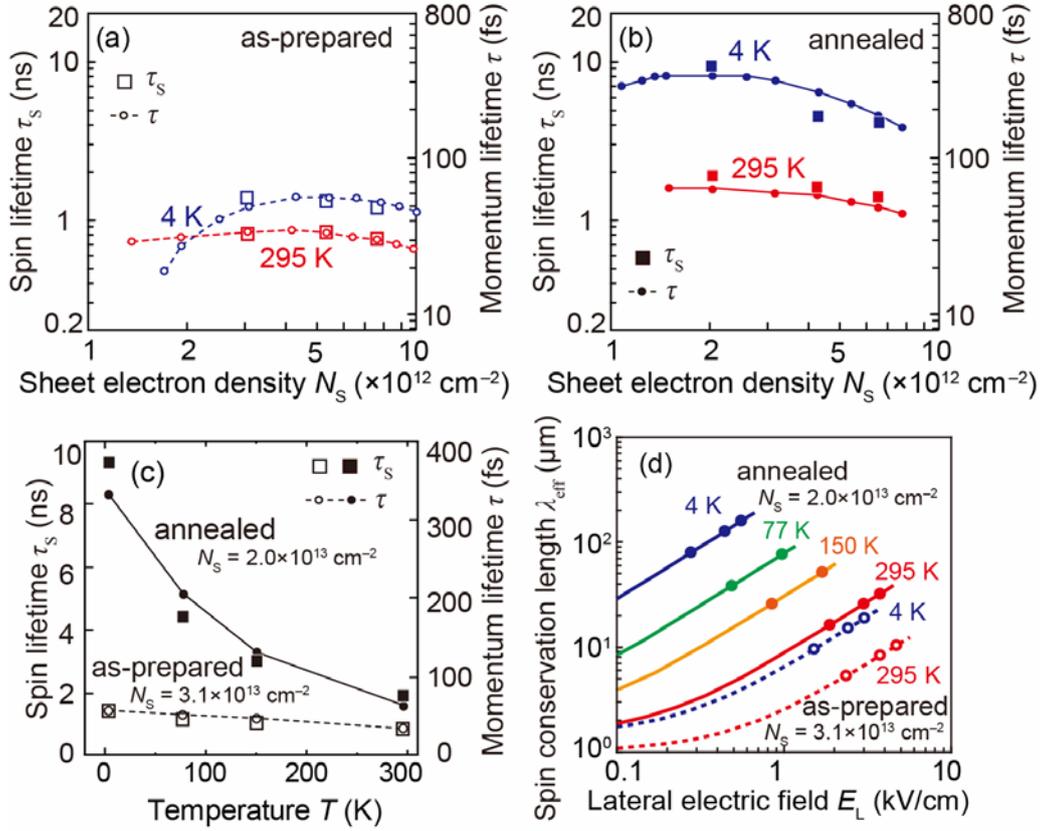

Figure 5 (a)(b) Spin lifetime $\tau_S$ and electron momentum lifetime $\tau$ plotted as a function of the sheet electron density $N_S$, which were estimated for the (a) as-prepared device and (b) annealed device. Left and right vertical axes represent the scales for $\tau_S$ and $\tau$, respectively. (a) Open squares are the $\tau_S$ values estimated from the 2TH signals and open circles with a dashed line are the $\tau$ values estimated from the electrical measurements (Fig. 2(a)). Blue and red colors denote $T = 4$ and 295 K, respectively. (b) Closed squares are the $\tau_S$ values estimated from the 2TH signals, and closed circles with a solid line are the $\tau$ values estimated from the electrical measurements (Fig. 2(b)). Blue and red colors denote $T = 4$ and 295 K, respectively. (c) $\tau_S$ and $\tau$ plotted as a function of $T$, where open circles/squares and closed circles/squares were estimated values for the as-prepared ($N_S = 3.1 \times 10^{12}$) and annealed devices ($N_S = 2.0 \times 10^{12}$), respectively. Left and right vertical axes represent the scales for $\tau_S$ and $\tau$, respectively. Squares are the $\tau_S$ values estimated from the 2TH signals and circles with a dashed or solid line are the $\tau$ values estimated from the electrical measurements (Fig. 2(a) and (b)). (d) Spin conservation length $\lambda_{eff}$ calculated using Eq. (5) plotted as a function of $E_L$, where open and closed circles are the experimental values estimated from the as-prepared and annealed devices, respectively, and dashed and solid lines are the extended curves calculated for the as-prepared and annealed devices, respectively. Blue, green, orange, and red colors denote $T = 4$, 77, 150, and 295 K, respectively.

Table 1 $N_S$ values used in the 2TH signal measurements for the as-prepared and annealed devices at various $T$.

| $T$ | $N_S$ (cm$^{-2}$) | | | |
|---|---|---|---|---|
| | 4 K | 77 K | 150 K | 295 K |
| as-prepared | $3.1\times10^{12}$ | $3.1\times10^{12}$ | $3.1\times10^{12}$ | $3.1\times10^{12}$ |
| | $5.3\times10^{12}$ | | | $5.3\times10^{12}$ |
| | $7.6\times10^{12}$ | $7.6\times10^{12}$ | $7.6\times10^{12}$ | $7.6\times10^{12}$ |
| annealed | $2.0\times10^{12}$ | $2.0\times10^{12}$ | $2.0\times10^{12}$ | $2.0\times10^{12}$ |
| | $4.3\times10^{12}$ | | | $4.3\times10^{12}$ |
| | $6.6\times10^{12}$ | $6.6\times10^{12}$ | $6.6\times10^{12}$ | $6.6\times10^{12}$ |

Supplemental Material

**Electron spin transport in a metal-oxide-semiconductor Si two-dimensional inversion channel: Effect of hydrogen annealing on spin scattering mechanism and spin lifetime**

Shoichi Sato[1,2], Masaaki Tanaka[1,2], and Ryosho Nakane[1]

[1]*Department of Electrical Engineering and Information Systems, The University of Tokyo, 7-3-1 Hongo, Bunkyo-ku, Tokyo 113-8656, Japan*

[2]*Center for Spintronics Research Network (CSRN), The University of Tokyo, 7-3-1 Hongo, Bunkyo-ku, Tokyo 113-8656, Japan*

## S1. Electrical properties of the 2D Si channel

We developed the procedure in this study with carefully considering the detailed electrical properties of Si 2D channels as described below. Owing to the quantum confinement in an inversion or accumulation channel, the 6-fold degenerated valleys at the conduction band minimum split into two subband groups (the 2-fold subband group with the lowest energy level and 4-fold subband group) and both the energy level and electron population in each subband can be changed with $E_G$ and $T$ [S1]. In addition to the electron momentum scattering processes in bulk Si materials, the Coulomb scattering and the surface roughness scattering take place due to the properties of the $SiO_2$ gate oxide and $SiO_2/Si$ interface. To gain a further insight into the temperature dependence of each scattering mechanism, we theoretically study how each momentum scattering mechanism changes with temperature. Here, the electron momentum lifetime due to each electron momentum scattering is referred to as follows: $\tau_{ac}$ due to intravalley acoustic phonon scattering, $\tau_f$ due to the intervalley optical phonon scattering through the *f* process, $\tau_g$ due to the intervalley optical phonon scattering through the *g* process, $\tau_C$ due to the Coulomb scattering, and $\tau_{sr}$ due to the surface roughness scattering. The total electron momentum lifetime $\tau$ is calculated by taking into account all of them through Matthiessen's rule.

We performed a self-consistent calculation of $\tau_{ac}$, $\tau_f$, $\tau_g$, $\tau_C$, and $\tau_{sr}$ in our 8-nm-thick Si inversion channel (see S.M. of refs. [7] and [8] for the detailed calculation process) using the following scattering parameters: the scattering parameter for the intervalley *f* or *g* process $D_f = D_g = 13 \times 10^8$ eV/cm, the root mean square (RMS) parameter of the interface roughness $\Delta\Lambda = 25 \times 10^{-8}$ m$^2$, and the fitting parameter for the Coulomb scattering $\alpha = 2.5 \times 10^{-9}$ cm$^4$/Vs (see Eq. S40(b) in S.M. of ref. [8]). These parameter values were determined so that $\tau$ reproduces the experimental electron mobility of the annealed sample at temperature $T$ (Figs.2(b) in the main manuscript). Figures S1(a) and (b) shows momentum lifetimes $\tau$, $\tau_{ac}$, $\tau_f$, $\tau_g$, $\tau_C$, and $\tau_{sr}$ (the contribution of each scattering mechanism) plotted as a function of $N_S$ at (a) 4 K and (b) 295 K, respectively  The Coulomb and surface roughness scatterings are dominant in the lower and higher $N_S$ ranges, respectively, whereas the phonon scatterings are negligible at low temperatures, but they become dominant at 295 K in the middle $N_S$ ($\sim 3 \times 10^{13}$ cm$^{-3}$) range. Figure S1(c) shows momentum lifetimes in the $T$ range from 4 to 295 K, where black, red, purple, orange, and blue curves represent $\tau$, $\tau_{ac}$, $\tau_f$, $\tau_g$, $\tau_C$, and $\tau_{sr}$, respectively. At low temperatures below 100 K, since temperature-independent $\tau_C$ is the lowest, the Coulomb scattering dominates $\tau$. (It is know that, unlike the case of a non-degenerated 3D electron gas, $\tau_C$ is insensitive to temperature in a degenerated 2D electron gas because the electron temperature is determined by the Fermi energy). As $T$ increases, $\tau_{ac}$ decreases gradually, whereas $\tau_f$ and $\tau_g$ decrease steeply due to the increase in the electron population at higher subbands. For more detail on $\tau_f$ and $\tau_g$, $\tau_f$ more steeply decreases than $\tau_g$ and $\tau_f$ is smaller than one-fourth of $\tau_g$ above 200 K. Among all the scattering processes, it is clear that the contribution of the intervalley $f$-process most largely increases as the temperature increases.

For further analysis, the contribution of each phonon scattering to the total phonon scattering are plotted in Fig S1(d), where red, orange, and purple regions represent the relative ratios of the contributions from $\tau_{ac}$, $\tau_g$, and $\tau_f$, respectively. The above-mentioned consideration is strongly supported by the fact that the relative ratio of $\tau_f$ increases with increasing temperature in Fig. S1(d).

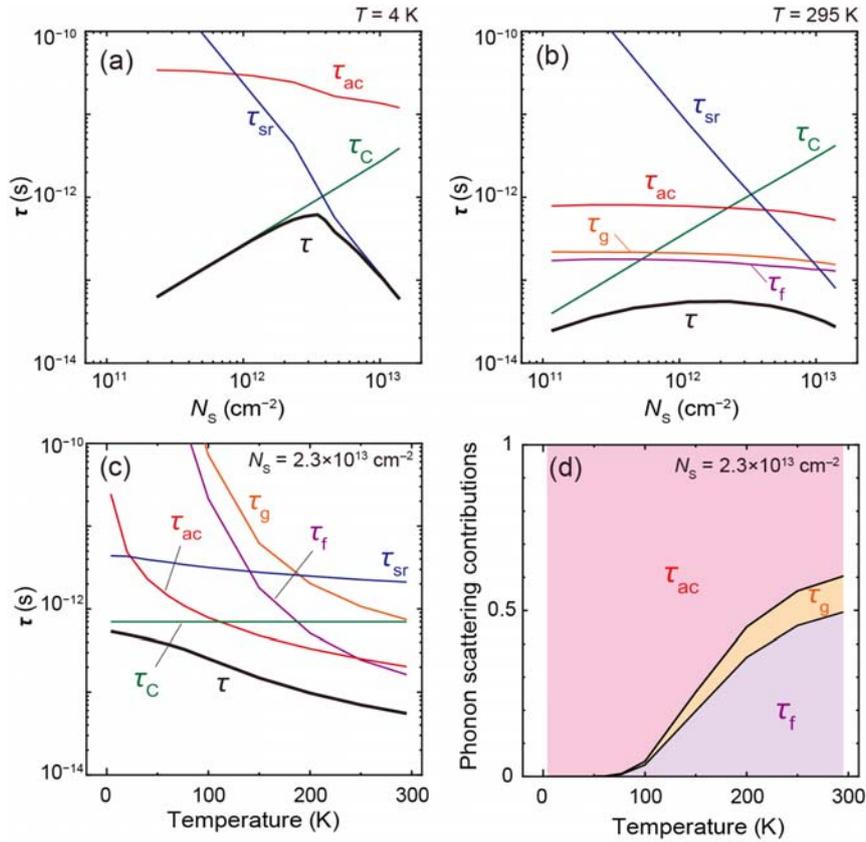

Figure S1 (a)(b)(c) Total momentum lifetime and the contribution of each scattering mechanisms plotted as a function of $N_S$ obtained by the SC calculation at (a) 4 K and (b) 295 K, and (c) those plotted as a function of $T$ at $N_S = 2.3\times10^{13}$ cm$^{-2}$. The black bold curve is $\tau$, and the blue, green, red, orange, and purple curves are $\tau_{ac}$ due to intravalley acoustic phonon scattering, $\tau_f$ due to the intervalley optical phonon scattering through the $f$ process, $\tau_g$ due to the intervalley optical phonon scattering through the $g$ process, $\tau_C$ due to the Coulomb scattering, and $\tau_{sr}$ due to the surface roughness scattering, respectively. (d) Contribution ratio of each phonon scattering processes to the total phonon scattering as a function of temperature. Red, orange, and purple regions represent the relative ratios of $\tau_{ac}$, $\tau_g$, and $\tau_f$, respectively.

## S2. Hallbar device and the Hall measurement

A bottom-gate-type Hall bar MOSFET having the same MOS structure as the spin MOSFET device examined in the main manuscript was prepared to characterize the electrical properties of the Si two-dimensional (2D) inversion channel; electron mobility $\mu$, momentum lifetime $\tau$, electron sheet density $N_S$, threshold voltage $V_{th}$, and the channel sheet resistance $R_S$. The device structure and measurement setup are shown in Fig. S2(a). The measurement procedure was the same as that of the spin MOSFET; first the as-prepared Hall bar devices were measured, annealed at 250°C for 30 min in a $N_2$+ 4% $H_2$ ambient, and then the annealed Hall bar devices were measured. Figure S3 shows $N_S$ estimated from the Hall voltage $V_T$ at 295 K as a function of $V_{GS}$, where open and closed circles are obtained in as-prepared and annealed Hall bar devices, respectively, and dashed and solid lines are the calculated using $N_S = C_{OX} (V_{GS} - V_{th})/q$, where $C_{OX}$ is the gate capacitance and $q$ is the elemental charge. We confirmed that these slopes agree with the designed value $C_{OX} = \varepsilon_{SiO2}/t_{BOX}$, where $\varepsilon_{SiO2}$ is the permittivity of $SiO_2$ and $t_{BOX}$ = 200 nm is the thickness of the buried oxide (BOX) layer. From the x-intercept of these lines, it was found that $V_{th}$ was negatively shifted from 13 V to 2 V by the annealing. Assuming that the shift in $V_{th}$ is caused only from the fixed charge at the $Si/SiO_2$ interface and in the BOX layer, the fixed charge density $N_{DEF}$ was estimated to be ~$1.5 \times 10^{12}$ cm$^{-2}$ and ~$2.3 \times 10^{11}$ cm$^{-2}$ for the as-prepared and annealed Hall bar device, respectively. This result indicates that the fixed charge was deactivated by the hydrogen atoms by the annealing [S3].

On the other hand, the channel sheet resistance $R_S$ was estimated from $V_L$ in Fig. S2(a). Note that $R_S$ value below ~200 K in the as-prepared device was estimated from the four-terminal measurement using the spin MOSFET device as shown in Fig. S2(b) because of the large gate leakage current in the as-prepared Hall bar device. Then, $\mu$ was estimated by $\mu = 1/qN_SR_S$ and plotted in Figs. 2(a) and (b) in the main manuscript.

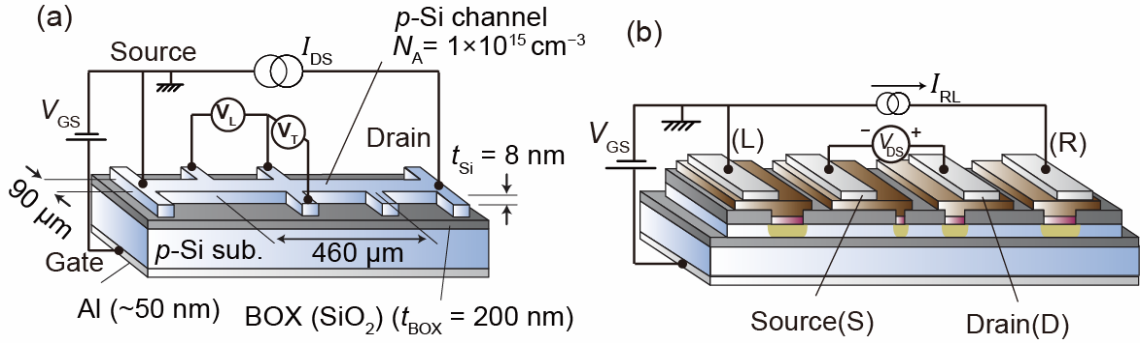

Fig. S2 (a) Schematic illustration of a Hall bar device with a 200-nm-thick SiO$_2$ buried oxide (BOX) layer. A 8-nm-thick p-Si channel with a Boron doping concentration $N_A = 1 \times 10^{15}$ cm$^{-3}$ has a length 460 μm and a width 90 μm. The measurement setup is also shown in the same figure, in which a constant bias current $I_{DS}$ and a constant source-gate voltage $V_{GS}$ were applied, and the longitudinal voltage $V_L$ and transverse voltage $V_T$ were measured while a magnetic field was applied perpendicular to the substrate plane. (b) Four-terminal measurement setup using the spin MOSFET device, where a constant current $I_{RL}$ is applied between the L and R electrodes, and the voltage $V_{SD}$ between the source (S) and drain (D) electrodes was measured.

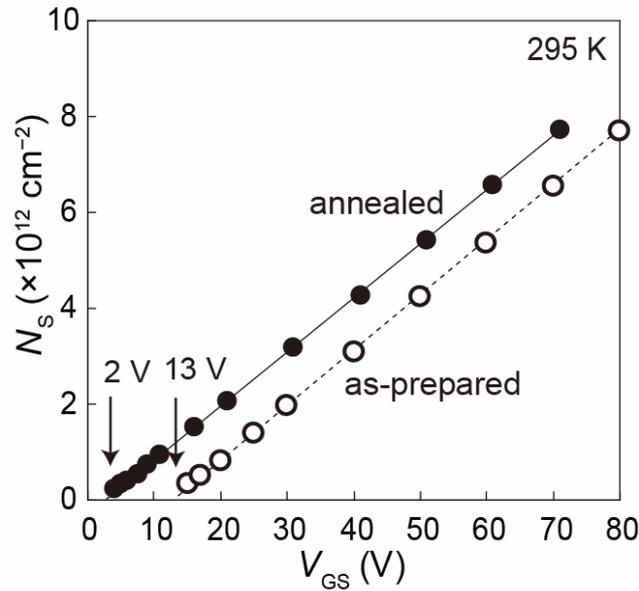

Figure S3 $N_S$ estimated from the Hall measurement as a function of $V_{GS}$, where open and closed circles are obtained in as-prepared and annealed Hall bar devices, respectively. Dashed and solid lines are calculated using $N_S = C_{OX} (V_{GS} - V_{th})/q$ with $C_{OX} = \varepsilon_{SiO2}/t_{BOX}$ and $V_{th}$ = 13 V and 2 V, respectively.

### S3. Enhancement factor of $\tau$, $\tau_S$ and $\lambda_{eff}$

The purpose here is to confirm the theoretical relation "spin conservation length in a high $E_L$ $\lambda_{\text{eff}}$ (~ $\lambda_{\text{drift}}$) is increased by $\alpha^2$ times when $\tau$ is increased by $\alpha$ times under a constant $\tau/\tau_S$", described in Section I in the main manuscript. To further analyze the enhancement of $\lambda_{\text{eff}}$ and $\tau_S$ by the annealing, these enhancement factors were obtained by $\beta = \lambda_{\text{eff}}^{\text{ANL}}/\lambda_{\text{eff}}^{\text{AS}}$ and $\zeta = \tau_S^{\text{ANL}}/\tau_S^{\text{AS}}$, respectively, where the superscripts AS and ANL denote the values for the as-prepared and annealed devices, respectively. Here, $\beta$ and $\zeta$ were calculated using the data in Fig. 5(c) and Eq. (5) with $E_L = 1$ kV/cm in the main manuscript. Figure S4 shows these enhancement factors plotted as a function of $T$, where open red and blue squares with dotted lines are $\beta$ and $\zeta$, respectively, and the enhancement factor of the momentum lifetime $\alpha = \tau^{\text{ANL}}/\tau^{\text{AS}}$ and $\alpha^2$ are also plotted by closed black circles with solid lines. In the entire $T$ range, $\beta$ and $\zeta$ reasonably agree with $\alpha^2$ and $\alpha$, respectively. Thus, the enhancement of $\tau_S$ is attributed to the enhancement of $\mu$, which is clear evidence for the above theoretical relation.

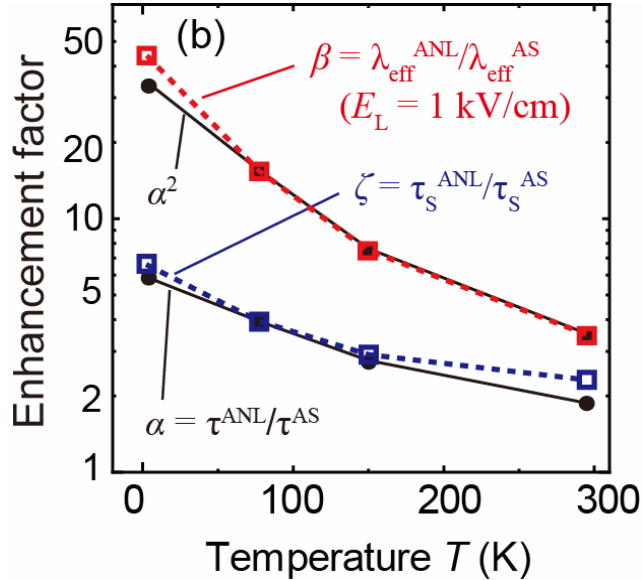

Figure S4 Enhancement factors $\alpha = \tau^{\text{ANL}}/\tau^{\text{AS}}$, $\alpha^2$, $\beta = \lambda_{\text{eff}}^{\text{ANL}}/\lambda_{\text{eff}}^{\text{AS}}$, and $\zeta = \tau_S^{\text{ANL}}/\tau_S^{\text{AS}}$ plotted as a function of $T$, where closed black circles with solid lines denote $\alpha$ and $\alpha^2$, and open red squares with a dashed line and open blue squares with a dashed line denote $\beta$ and $\zeta$, respectively. These values were calculated using the data in Fig. 5(c) and Eq. (5) with $E_L = 1$ kV/cm in the main manuscript.